\documentclass[]{interact}

\usepackage{epstopdf}
\usepackage[caption=false]{subfig}

\usepackage[numbers,sort&compress]{natbib}
\bibpunct[, ]{[}{]}{,}{n}{,}{,}
\renewcommand\bibfont{\fontsize{10}{12}\selectfont}

\theoremstyle{plain}

\theoremstyle{definition}

\theoremstyle{remark}

\usepackage{hyperref}

\title{Assessing adult physical activity and compliance with 2008 CDC guidelines using a Bayesian two-part measurement error model \footnote{This is an Accepted Manuscript of an article published by Taylor \& Francis in Journal of Applied Statistics on June 21, 2022, available at: \url{https://doi.org/10.1080/02664763.2022.2088706}}}

\author{
\name{Daniel Ries\textsuperscript{a}\thanks{CONTACT D.~Ries. Email: dries@sandia.gov} \footnote{} and Alicia Carriquiry\textsuperscript{b}}
\affil{\textsuperscript{a}Statistics and Data Analytics Department, Sandia National Laboratories, Albuquerque, NM} 
\affil{\textsuperscript{b}Department of Statistics, Iowa State University, Ames, IA} 
}

\begin{document}

\maketitle

\begin{abstract}
While there is wide agreement that physical activity is an important component of a healthy lifestyle, it is unclear how many people adhere to public health recommendations on physical activity.  The {\it Physical Activity Guidelines (PAG)}, published by the CDC, provide guidelines to American adults, but it is difficult to assess compliance with these guidelines. The PAG further complicate adherence assessment by recommending activity to occur in at least 10 minute \emph{bouts}.  To better understand the measurement capabilities of various instruments to quantify activity, and to propose an approach to evaluate activity relative to the PAG, researchers at Iowa State University administered the Physical Activity Measurement Survey (PAMS) to over 1,000 participants in four different Iowa counties.  In this paper, we develop a two-part Bayesian measurement error model and apply it to the PAMS data in order to assess compliance to the PAG in the Iowa adult population. The model accurately accounts for the 10 minute \emph{bout} requirement put forth in the PAG. The measurement error model corrects biased estimates and accounts for day to day variation in activity. The model is also applied to the nationally representative National Health and Nutrition Examination Survey. 

\end{abstract}

\begin{keywords}
measurement error; two-part model; moderate to vigorous physical activity; Bayesian; NHANES
\end{keywords}

\section{Introduction}
Physical activity has an undisputed role in a healthy lifestyle. Regular physical activity has been linked to prevention and treatment of cardiovascular disease \cite{ warburton,reiner}, diabetes \cite{warburton,reiner}, cancer \cite{warburton}, hypertension \cite{warburton}, osteoporosis \cite{warburton}, depression \cite{ warburton}, obesity \cite{warburton, reiner}, and Alzheimer's disease \cite{reiner}.

In recent years the pace of research in physical activity and its effect on health has accelerated.  According to the Centers for Disease Control and Prevention (CDC), over 70\% of Americans age 20 and over are overweight or obese, and almost 40\% are obese  \footnote{{https://www.cdc.gov/nchs/fastats/obesity-overweight.htm}}. 
In 2008, the Department of Health and Human Services issued the \emph{2008 Physical Activity Guidelines for Americans} (PAG)\footnote{{https://health.gov/paguidelines/guidelines/}}.
This was the first time that physical activity guidelines were published by the federal government. 

The PAG recommends that adults spend at least 150 minutes each week carrying out moderate-intensity activity, at least 75 minutes in vigorous-intensity activity, or some combination of moderate to vigorous physical activity (MVPA). Furthermore, it recommends that this activity be in intervals, or {\it bouts}, of at least 10 minutes. They define moderate-intensity to be a 5 or 6 on an intensity scale of 0 to 10; a brisk walk is an example. Vigorous-intensity is a 7 or 8 on the same scale; jogging or lap swimming are examples. In addition, the PAG recommends doing muscle-strengthening activities that involve all major muscle groups twice or more per week. These are the minimum levels of activity that are expected to have an effect on health. The report also advises that any physical activity above the minimum will result in additional health benefits. 

A public health question is: what proportion of  the population adheres to these guidelines? And does that proportion change based on age, sex, or other demographic variables?  This type of information is important for policy makers not only to assess compliance but also to design interventions that target certain subpopulations. Yet, there is no agreement about how to measure physical activity. Furthermore, the only nationally representative source of physical activity measurements is the National Health and Nutrition Examination Survey (NHANES) \footnote{https://wwwn.cdc.gov/nchs/nhanes/default.aspx}. To better understand the measurement error associated with different instruments, Iowa State University conducted the NIH-funded Physical Activity Measurement Survey (PAMS) to collect physical activity information \cite{beyler}.  The objectives of the study were to understand the measurement error of different methods of measuring physical activity in adults.  The PAMS data helps develop a Bayesian two-part modeling approach that accounts for measurement error in physical activity observations. That modeling approach can be used to determine the proportion of the Iowa adult population which adhere to the PAG. Results are compared at a national level with NHANES data and possible reasons for the differences are discussed.

This paper is organized as follows.  Section 2 introduces PAMS and reviews the literature on approaches to measure physical activity.  Section 3 develops a two-part model to jointly model the distribution of daily 10-minute bouts and the average excess minutes of MVPA. Section 4 presents the fitted model's results in the context of the application.  In this section, we also compare results with those obtained using NHANES data and discuss differences. Section 5 presents model diagnostics and goodness of fit.   Section 6  discusses results and suggests future work.

\section{The Physical Activity Measurement Survey (PAMS)} \label{pams}

\subsection{A brief review of physical activity measurement}

The term ``physical activity'' is not well defined. It is hardly surprising that many different approaches to quantify physical activity have been proposed in recent years.  Thinking of physical activity as the amount of energy expended per day during a short period (e.g., two weeks), then doubly-labeled water is considered to be the gold standard among measurement instruments \cite{bouten, hall11}.  However, it is impractical to use doubly-labeled water in large studies, not only because of cost but also because of respondent burden.

In practice, instruments such as accelerometers that measure movement have become common-place.  Accelerometers provide estimates of movement through uni-,bi-, or triaxial measurements. Measurements of activity are then often reported as ``counts" such as with the Actigraph or metabolic equivalents (MET) for Sensewear Armbands (SWA). Typically, raw accelerometer data are converted to counts or METs using proprietary algorithms, but there are new attempts to analyze raw accelerometry data \cite{bai}. Urbanek et al. \cite{urbanek} uses the full, raw accelerometry data to create new measures of stride-to-stride gait variation. He et al. \cite{he} uses wavelets \cite{bai2}  to classify activity types based on accelerometry data. This may help compensate for the fact that accelerometry data provide no information about {\it the context} in which physical activity takes place. There is a rich literature that focuses on the relationship between total counts per day and some outcome variable \cite{schrack,steeves}. Other authors use count data at the hour level to further understand how physical activity levels vary by demographic groups \cite{steeves, steeves2}. Functional data analysis is also a way to model and analyze high-frequency, accelerometer data and the short term timeframe such as how varying activity levels within a day affect covariates. Xiao et al. \cite{xiao} provides methods to model the systematic and random patterns of physical activity while accounting for dependence on covariates such as age and gender. Fan et al. \cite{fan} using functional ANOVA to assess the circadian activity profiles of teenage girls. Goldsmith et al. \cite{goldsmith} uses functional scalar regression to understand the association between physical activity and a variety of covariates.

\subsection{Description of PAMS}

The PAMS was conducted over two years starting in 2009 in Iowa. The goal was to obtain information on physical activity of adult men and women. The survey was conducted in two stages across four counties and included two strata per county. In each county there was a ``high minority population" and ``low minority population" stratum to improve chances of recruiting African American and Hispanic individuals. Eligible participants included adults between 21 and 71, with the ability to engage in physical activity, who were not pregnant or lactating, were able to speak English or Spanish, and had a landline in their  place of residence.  A summary of the demographic characteristics of PAMS participants is given in Table \ref{pamsdemographics}.

\begin{table}[h]
\centering
\caption{Demographic characteristics  of PAMS. BMI indicates Body Mass Index.}
	\begin{tabular}{|l|l|l|}
	\hline
	& Female &  Male \\
	\hline
	Count & 630 & 427 \\
	\hline
	Age 21-39 & 106 & 129 \\
	Age 40-59 & 331 & 187 \\
	Age 60-71 & 193 & 111 \\
	\hline
	BMI Range & 16.8-72.9 & 17.5-60.3 \\
	Mean BMI & 30.87 & 30.1 \\
	Standard Deviation BMI & 6.27 & 8.09 \\
	\hline
	African Americans Count & 54 & 30 \\
	Hispanics Count & 20 & 16 \\
	\hline
	Smokers Count & 106 & 83 \\	
	\hline
	 Graduates Count & 229 & 182 \\
	\hline
	Physical Jobs Count & 287 & 215 \\
	\hline
	\end{tabular}
\label{pamsdemographics}
\end{table} 

Energy expenditure (EE) information was collected on two separate occasions using SWA. In order to mitigate dependence in activity across days for an individual, the two measurements were taken 2-3 weeks apart. The SWA provides MET levels every minute. A MET is a measure of energy cost for a particular physical activity. Formally, 1 MET is defined as 0.0175 kcal/kg/min expended. METs can be thought of as a multiplicative effort to carry out the activity relative to resting state. An activity that is classified as 5 METs then requires about 5 times the energy that is required to be at rest. MET-minutes are the number of minutes in an activity multiplied by the MET value of that activity (we include a figure in the supplemental material of the raw data for three individuals from the SWA for reference).

The method with which the SWA calculates MET-minutes is proprietary, but SWA's measurement properties and validity have been studied.  Hills et al. \cite{hills} and \cite{hill} found the SWA to be an accurate measure of physical activity. Santos et al. \cite{santos} and  Scheers et al. \cite{scheers} found that the SWA tends to overestimate MVPA. Calabro et al. \cite{calabro} also found the SWA to slightly overestimate physical activity, but it was much closer to truth than the other accelerometers used which underestimated physical activity by a greater magnitude. Casiraghi et al.  \cite{casiraghi} notes that the SWA is a good measurement for certain activities  like running and walking, but its use is limited in activities like cycling and swimming.   The \emph{Compendium of Physical Activities}\footnote{https://sites.google.com/site/compendiumofphysicalactivities/home}  gives MET values for many common daily activities.

The PAG defines activities with METs ranging from 3.0 to 6.0 as moderate intensity and activity with METs greater than 6.0 as vigorous. This means that the recommendation of 150 minutes of moderate physical activity is equivalent to 150$\times$3.0 = 450 MET-minutes per week.  Another stipulation in the \emph{PAG} is that activity must occur in bouts of at least 10 minutes to count toward this total. In practice, what constitutes a  \emph{bout} is less clear. 
To address the research questions  we  need an operational definition of what constitutes a  \emph{bout}. 

\subsection{Definition of  \emph{Bouts} and Average Excess MET-minutes}

We define a  \emph{bout} as a burst of activity in at least 8 out of 10 minutes, in at least 3 METs \cite{tucker11, kim}, as a bout. This means that at least 8 minutes out of 10 minutes must be in at least moderate physical activity to count toward the recommended guidelines. We allow the 8 out of 10 to move along a rolling window, by shifting a 10 minute window, minute by minute, to determine if the time in the moving window counts towards at least 3 MET activity. As long as we observe $\leq 2$ minutes of less than moderate activity ($<$3 METs), the clock continues to count MET-minutes for that bout. Once there are $\geq 3$ minutes in less than moderate activity, the ``clock stops" at the minute before the 3rd minute is reached. Further, the final 2 minutes of activity cannot be below moderate level. 
About 24\% of the 24 hour data collections had zero bouts, and 11\% of individuals in PAMS had zero bouts on both study days. 

 Total MET-mins in MVPA is zero for individuals with zero bouts, and is $\geq$30 for individuals with a minimum of one bout  (10 minutes $\times$ 3 METs = 30 MET-mins in MVPA). To account for these constraints, define the random variables as:
\begin{align}
 Y_{1ij} & \text{: Number of bouts for individual $i$ during day $j$} \\
 Y_{2ij} & \text{: Average MET-mins per bout - 30, set as zero if $Y_{1ij}=0$}.
\end{align}

\begin{figure}
\centering
\includegraphics[height=2.5in,width=4in]{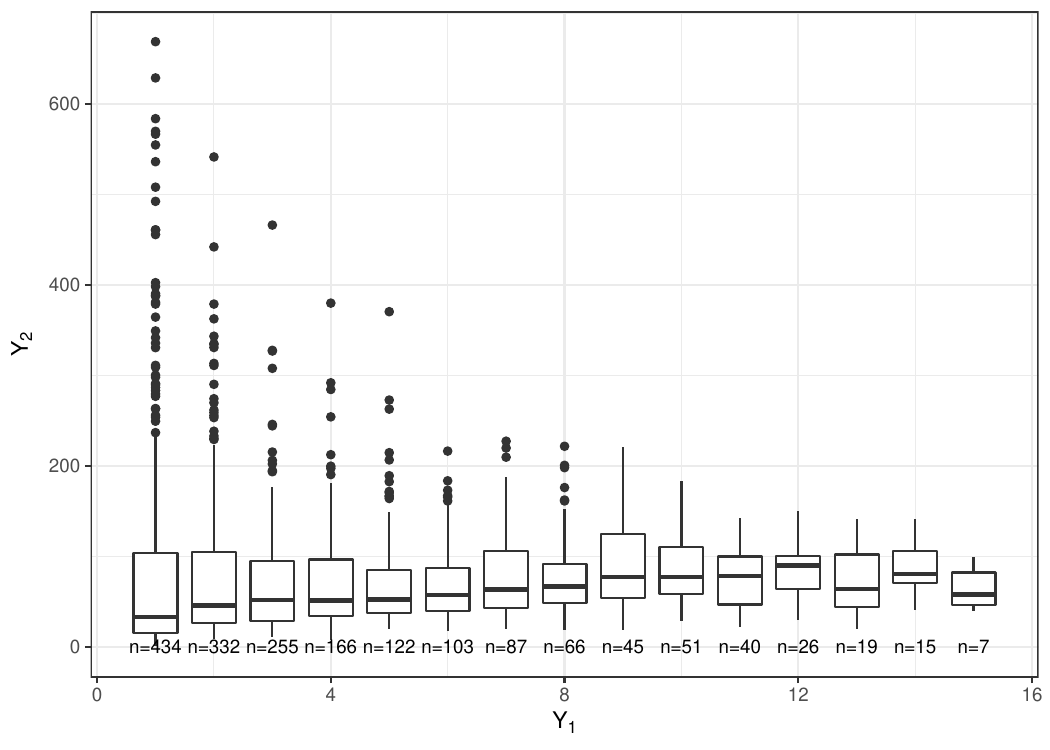}
\caption{$Y_2$ vs $Y_1$ in 24 hour period for all individuals and days.}
\label{y1vsy2}
\end{figure}

Denote ${\bf Y}_{ij} = (Y_{1ij},Y_{2ij})$, ${\bf Y}_1=\{Y_{1ij}\}_{\forall i,j}$, ${\bf Y}_2=\{Y_{2ij}\}_{\forall i,j}$, and ${\bf Y} = ({\bf Y}_1,{\bf Y}_2) $. Refer to $Y_{2ij}$ as the \emph{average excess MET-minutes}. 
There were several outliers in both number of bouts and total MET-minutes; we removed  persons with more than 2500 total MET-minutes because they are believed to be mismeasurements.  Figure \ref{y1vsy2} plots $Y_2$ against $Y_1$. The range in the plot is constrained between 1 and 13 per day to ensure that that there are at least 20 observations in each bout boxplot.  Even after accounting for the number of bouts, the medians of $Y_{2ij}$ is positively related to number of bouts. This means those who have more bouts, often engage in longer or more intense bouts. 
Additionally, the distribution of residuals of $Y_2$ is right skewed, and a log transformation makes these residuals resemble a Normal distribution. Further details and plot are in Section 1 of the supplemental material.

\subsection{Checking for Day Effect of Observations}

Creating a two way contingency table for number of bouts in day one versus number of bouts in day two allows for checking whether exchangeability is a reasonable  assumption for $Y_{i1}$ and $Y_{i2}$ using Bowker's test \cite{bowker}. The p-value for the hypothesis test was 0.12 indicating exchangeability is not an unreasonable assumption. The contingency table and further description of the test is Section 2 of the supplemental material. We also checked for weekend effect on $Y_1$ using a paired t-test, which resulted in a p-value  $=$0.50. Since there is no obvious indication of a day effect,  bouts within individuals are assumed exchangeable. 

A day effect is also possible for $Y_2$, which depends on the number of bouts (Figure \ref{y1vsy2}). There is also interest in knowing whether there is an effect of weekend on $Y_2$. To explore the association between $Y_2$ across days, we  fit the following linear model:

\begin{align}
	\begin{split}
	Y_{2i1}-Y_{2i2} &= \beta_0 + \beta_1 (Y_{1i1}-Y_{1i2}) + \beta_2 (Weekend_{i1}-Weekend_{i2}) + \epsilon_{i}, i=1,2,...,n \\
	\epsilon_i &\overset{iid}{\sim} N(0,\sigma^2),
\label{y2exchreg}
\end{split}
\end{align}

where $Weekend_{ij}$ is an indicator for weekday (M-F) versus weekend (Sat or Sun),  $\beta_0$ represents the day effect and $\beta_2$ represents weekend effect. Hypothesis tests for day or a weekend effect on $Y_2$ indicate no effect (p-value = 0.63, 0.65, respectively). We also checked for weekend effect on $Y_1$ using a paired t-test, which resulted in a p-value  $=$0.50. Since there is no obvious indication of a day effect, we will assume that average excess MET-minutes within individuals are exchangeable.

\section{Model for MET-mins in MVPA During at least 10 Minute Bouts}

We introduced the correlated random variables $Y_{1ij}$ and $Y_{2ij}$ earlier as the response variables. In this section, we present a measurement error model for $Y_1$ and $Y_2$.

\subsection{Notation and Data}
After removing outliers and individuals without a replicate observation, we have $N=2114$ observations obtained on $n=1057$ individuals. We let $i$ represent individual, $i=1,...,1057$ and $j$ represent the measurement occasion, $j=1,2$. We define a vector ${\bf Z}_i$  of dimension eight, that includes covariates for individual $i$:  gender, age, indicators for Black, Hispanic, smoker, college degree, and physical job. The full model matrix is ${\bf Z}=({\bf Z}_1,{\bf Z}_2,...,{\bf Z}_{1057})'$.  There were 315 instances of item non-response for occupation in the 1057 individuals, so we imputed the missing values using predictions from a logistic regression with physical job as the response and all remaining covariates in ${\bf Z}$ as covariates. Denote $T_{1ij}$  as individual $i$'s unobservable  true number of bouts on day $j$ and $T_{2ij}$ as individual $i$'s unobservable true average  excess MET-minutes per bout on day $j$. We let $t_{1i}$  and $t_{2i}$ be the expected values of $T_{1ij}$ and $T_{2ij}$ conditional on individual $i$, respectively. We refer to these quantities as individual $i$'s \emph{usual} number of bouts in a day and \emph{usual} average excess MET-mins per bout, respectively. More formally:
\begin{align}
\begin{split}
	t_{1i} &\equiv E(T_{1ij}|i), \\
	t_{2i} &\equiv E(T_{2ij}|i).
\label{usualdef}
\end{split}
\end{align}

Following \cite{kipniszeros1} we assume that the measurements of physical activity are unbiased for the usual activity levels. This is  a plausible assumption because the measurements are obtained using an objective instrument. We recognize that this is a strong assumption since we cannot validate it with the data we have. Future work should design data collections, such as done in \cite{shook}, in order to properly assess or mitigate this assumption. 
We also assume that the armband records zero bouts if and only if individual $i$ participated in zero bouts of activity on day $j$. Formally, these assumptions can be expressed as:

\begin{align}
\begin{split}
	t_{1i} &= E(Y_{1ij}|i), \\
	t_{2i} &= E(Y_{2ij}| Y_{2ij} >0, i) \times P(Y_{2ij} > 0|i),  \\
	P(T_{1ij} > 0|i) &= P(Y_{1ij} > 0|i).
\label{notation1}
\end{split}
\end{align}

By construction, $ P(Y_{2ij} > 0|i)= P(Y_{1ij} > 0|i)$. To answer the original question of adherence to the PAG,  individual $i$'s  \emph{usual} total MET-minutes in MVPA for a day is defined as:

\begin{align}
	t_{3i} &\equiv 30 t_{1i} + t_{2i} \times t_{1i}.
\label{notation2}
\end{align}

\subsection{Modeling Number of Bouts}

Individual $i$'s number of bouts at measurement $j$, $Y_{ij}$ is a count, so a natural model is the Poisson distribution. However, as Figure 3 in the supplemental material shows, there is within person overdispersion present, so a standard Poisson model is not flexible enough for the PAMS data. We also considered a Negative Binomial distribution to handle the overdispersion. During model assessment, the Generalized Poisson proved to be a better fit (see Section \ref{sec:assessment}).

An alternative to the Poisson distribution that allows for a more flexible mean-variance relationship is the Generalized Poisson distribution \cite{consul}. The Generalized Poisson distribution is indexed by two parameters, $\theta$ and $\lambda$, with probability density function

\begin{align}
f(x|\theta,\lambda) &= \theta (\theta + x\lambda)^{x-1}\frac{e^{-(\theta + x\lambda)}}{x!}, x=0,1,2,... 
\end{align}

The Generalized Poisson is overdispersed relative to a Poisson distribution if $\lambda > 0$, underdispersed if $\lambda < 0$ and a regular Poisson if $\lambda=0$.  
When $0 < \lambda < 1$, the probability mass function and first two moments of the distribution can be written directly without truncation or normalization \cite{consul, scollnik}. In this case, its expected value is $\frac{\theta}{1-\lambda}$ and the variance is $\frac{\theta}{(1-\lambda)^3}$. Reparametrizing the distribution in terms of the mean, $\mu$, the variance is $\frac{\mu}{(1-\lambda)^2}$. 
At this point we only concern ourselves with overdispersion, thus the restriction that $0 < \lambda < 1$ is appropriate. 

We model the mean of the Generalized Poisson distribution as a function of the covariates plus an individual random effect for across person overdispersion. The random effects are assumed to be joint Normal with random effects with average excess MET-minutes. The priors for $\lambda$ and $\boldsymbol{\gamma}$ are proper and independent, but relatively non-informative. The model for $Y_{1ij}$ is written as:

\begin{align}
\begin{split}
	Y_{1ij} | b_{1i}, Z_i &\overset{ind}{\sim} \text{GenPoisson}(\mu_{1i},\lambda) \\
	\mu_{1i} &= E(Y_{1ij}|i) = e^{Z_i' \boldsymbol{\gamma}+b_{1i}} \\
	\lambda &\sim \text{Uniform}(0,1) \\
	\boldsymbol{\gamma} &\sim N \left( {\bf 0}_8, 100 I_{8 \times 8} \right).
\label{y1model}
\end{split}
\end{align}

\subsection{Modeling Average Excess MET-minutes} 

Average Excess MET-minutes can take positive value or be zero; this type of data is commonly referred to as ``semicontinuous data" and occurs often in the fields of epidemiology and nutrition. Many models for semicontinuous data have built upon the work of \cite{olsen, tooze, nci}. \cite{neelonbayes} and \cite{neelonspatial} propose Bayesian approaches for estimation in these models.   
Kipnis et al. \cite{kipniszeros1} and \cite{kipniszeros2} propose a measurement error approach for semicontinuous data via regression calibration in the context of a   nutrition application.

To account for measurement error and the large number of zeros in the sample, we propose the following model for total excess MET-minutes:

\begin{align}
\begin{split}
	Y_{2ij}|b_{2i}, Z_{1i} &\overset{ind}{\sim} (1-\pi_{i}) I(Y_{2ij}=0) + \pi_{i} \text{LogNormal}(\mu_{2i},\sigma_y^2)I(Y_{2ij}>0), i=1,...,n, j=1,2, \\
	\mu_{2i} &= E(\log Y_{2ij}|Y_{2ij}>0,i)  = {Z_i' \boldsymbol{\beta} + b_{2i}}, \\
b_{1i},b_{2i} &\overset{ind}{\sim} N \left(
\begin{bmatrix}
0 \\
0
\end{bmatrix} , \Sigma_b \right), \\
	\Sigma_b &\sim \text{Inverse-Wishart}(3, I_{2 \times 2}) \\
	\sigma_y^2 &\sim \text{Inverse-Gamma}(0.01,0.01), \\
	\boldsymbol{\beta} &\sim N \left( {\bf 0}_8, 100 I_{8 \times 8} \right),
\label{y2model}
\end{split}
\end{align}

where $\pi_i= P(t_{2ij} > 0|i) = P(Y_{1ij} > 0|i)$ is individual $i$'s probability of participating in at least one bout, which can be calculated using the Generalized Poisson probability mass function given parameters $\boldsymbol{\gamma},\lambda,b_{1i}$ and covariates ${\bf Z}_i$. The priors for $\boldsymbol{\beta},\sigma_y^2,\Sigma_b$ are conjugate, independent, and relatively noninformative. Sensitivity analysis showed little effect of the priors on inference for the variance components.

The full likelihood for an individual can be written as:

\begin{align}
L_i(\boldsymbol{\theta}) &= \int \int \prod_{j=1}^{2} f(Y_{2ij}|\boldsymbol{\theta},Z_i,b_{2i},b_{1i}) f(Y_{1ij}|\boldsymbol{\theta},Z_i,b_{1i}) f(b_{1i},b_{2i}|\boldsymbol{\theta}) db_{1i} db_{2i},
\end{align}

where $f(Y_{1ij}|\cdot)$, $f(Y_{2ij}|\cdot)$, and $f(b_{1i},b_{2i}|\cdot)$ are as defined in Equations \eqref{y1model} and \eqref{y2model}, and $\boldsymbol{\theta}$ is a vector of all unknown parameters. Along with the assumption of independence between individuals, the full likelihood is:

\begin{align}
 L(\boldsymbol{\theta}) &= \prod_{i=1}^{n}  L_i(\boldsymbol{\theta}).
\end{align} 

\subsection{Estimating Distribution of Usual Daily MVPA}
\label{sec:usual_mvpa}
Our goal is to estimate the proportion of Iowans who are in compliance with the PAG on average. To answer this question, we focus on the distribution of \emph{usual} total MET-minutes  in MVPA for individuals from a specified population in a day. We specify the population in which we are interested  through the design matrix ${\bf Z}$. To estimate this distribution,  simulate draws of $t_3$ through the following:

For $\ell$ from $\ell=1,2,...,L$ do:
\begin{enumerate}
\item
Sample $\boldsymbol{\theta}^{(\ell)}$ from the posterior distribution $p(\boldsymbol{\theta}|{\bf Y})$.
\item
Simulate ${ b_{1i}^{(\ell)}}$ and ${ b_{2i}^{(\ell)}}$ from $p(b_{1i},b_{2i}|\boldsymbol{\theta}^{(\ell)},{\bf Z}_i)$ for $i=1,...,n$.
\item
Compute $t_{3i}^{(\ell)}= 30 \times E(Y_{1ij}|\boldsymbol{\theta}^{(\ell)},{\bf Z}_i,b_{1i}^{(\ell)})$ + $E(Y_{2ij}|Y_{2ij}>0,\boldsymbol{\theta}^{(\ell)},{\bf Z}_i,b_{2i}^{(\ell)}) $ $\times$  $P(Y_{1ij}>0|\boldsymbol{\theta}^{(\ell)},{\bf Z}_i,b_{1i}^{(\ell)}) \times E(Y_{1ij}|\boldsymbol{\theta}^{(\ell)},{\bf Z}_i,b_{1i}^{(\ell)}))$ for $i=1,...,n$, as defined in Equation \eqref{y1model} and \eqref{y2model}.
\end{enumerate}

The proportion of individuals from the  population who meet the PAG in the $\ell^{th}$ draw is given by:

\begin{align}
 p^{(\ell)} &= \frac{1}{n} \sum_{i=1}^{n} I \left( t_{3i}^{(\ell)} \geq \frac{450}{7} \right) .
\end{align}

If there are weights $w_i$ associated with the individuals of the design matrix ${\bf Z}$, estimates of percentiles of the distribution of $t_3$ can be obtained by:

\begin{align}
  p^{(\ell)}_{weighted} &= \frac{1}{\sum_{i=1}^{n}w_i} \sum_{i=1}^{n} I \left( t_{3i}^{(\ell)} \geq \frac{450}{7} \right)w_i .
\end{align}

Recall that our model is estimating usual \emph{daily} MET-minutes in bouts, and our model already considers how often individuals participate in at least a bout of MVPA. Because of this, we can consider weekly activity to be 7$\times$usual daily MET-minutes in bouts.

\section{Results}

We proceed with estimation via MCMC following \cite{neelonbayes} and \cite{neelonspatial}, who propose Gibbs algorithms for two-part models with semicontinuous data that are nearly or completely conjugate. We construct a Gibbs algorithm for drawing samples from the posterior distribution, and since many of the priors are not conjugate, we need to use a Metropolis-within-Gibbs sampler. 
The Gibbs algorithm was written in C++ and R. Full conditional distributions can be found in Section 4 of the supplemental material. Starting values for the MCMC are obtained from maximum likelihood. We used the resulting MLE's and lower and upper bound of 99.99\% confidence intervals as well dispersed starting values for the regression parameters in order to use the Gelman-Rubin diagnostic in assessing chain convergence. We dispersed $\lambda$ between 0 and 1 for its starting values in the 3 chains. Values for $\Sigma_{b}, \sigma_y^2$  were chosen such that starting values were far above and below the final region of the posterior distribution. 

We ran 3 chains of length 500,000, with the first 50,000 draws as burn-in, and thinned every 15 iterations to save on memory and reduce the autocorrelation of parameter draws. Traceplots and Gelman-Rubin diagnostics (all  $<$ 1.05)  indicated good mixing and no signs of non-convergence. The Monte Carlo standard error was calculated using the R package \texttt{mcmcse}. The MC error was less than 1.5\% of posterior standard deviation for all parameters. 

\subsection{Parameter Estimates}

\begin{figure}
\includegraphics[width=0.8\textwidth]{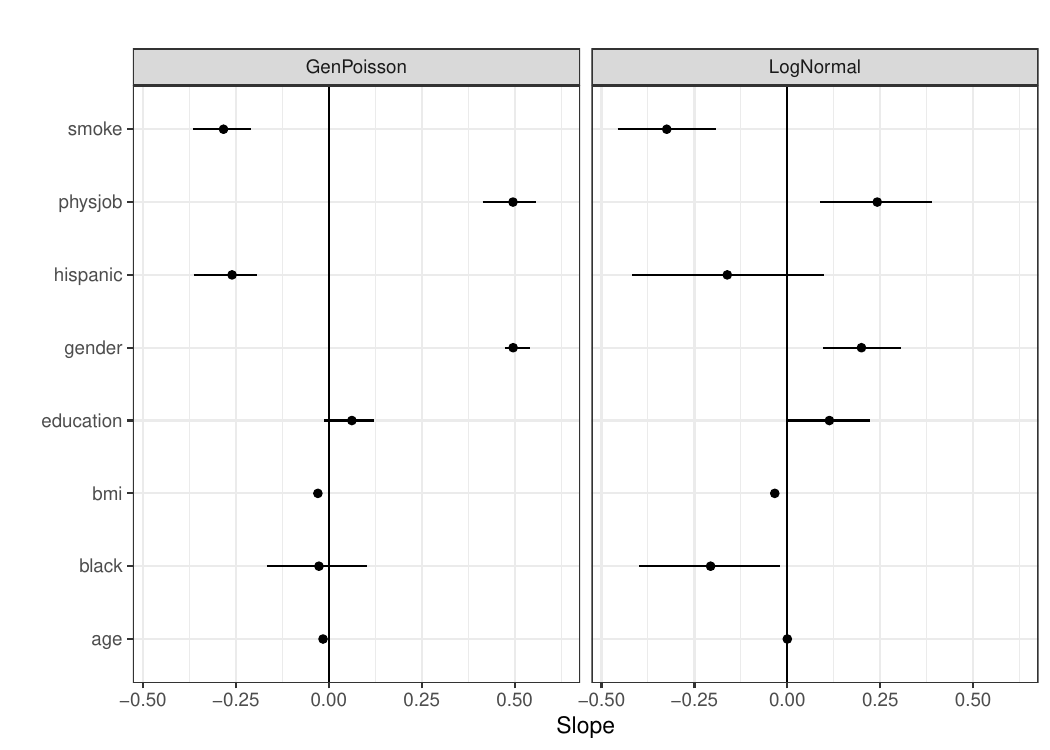}
\caption{Posterior means and 95\% credible intervals for regression coefficients for both parts of model.}
\label{regcoefs}
\end{figure}

Figure \ref{regcoefs} shows posterior means and 95\% credible intervals for all regression coefficients in the model. The signs on the coefficients and relative interval widths nearly match for all covariates across the two parts of the model. Males and those with physical jobs tend to exhibit a higher number of bouts per day and more average excess MET-minutes per bout. BMI is negatively associated with bouts per day as well as average excess MET-minutes per bout.  Age is negatively associated with  bouts but not with average excess MET-minutes per bout. Having a college education was positively associated with average excess MET-minutes per bout, but not with number of bouts. Hispanic was negatively associated with bouts per day. Black is negatively associated with average excess MET-minutes but there is considerable uncertainty. Smoking is negatively associated with both number of bouts and average excess MET-minutes.

Physical jobs being positively associated with bouts and average excess MET-minutes makes intuitive sense, as people with these jobs are engaging in physical activity throughout the workday, and because men more often have these jobs. The negative associations with BMI could be explained by those who participate in activity are less likely to be overweight since those individuals are seeing the benefits of physical activity. Those with a college education may be more likely to have non-physical jobs, so it is possible they get their physical activity through voluntary exercise. This exercise could happen all at once outside work hours, which would explain the non-relationship with number of bouts. 

Table \ref{estimates} shows posterior means and 95\% credible intervals for the remaining parameters. Recall that in a Generalized Poisson distribution, a value of $\lambda > 0$ indicates overdispersion. For the PAMS data, the estimate of $\lambda$ was 0.09 (0.08,0.1). The estimated measurement error variances, $\sigma_{b_1}^2,\sigma_{b_2}^2$, are  large relative to the regression coefficients corresponding to their respective model component. This suggests that there is considerable day to day variation in physical activity and that the device measurements themselves are noisy. The estimate of $\rho_b$ is 0.41 (0.18,0.77), indicating that there is a significant amount of correlation between the mean functions of $Y_1$ and $Y_2$.  

\begin{table}[ht]
\centering
\caption{Posterior mean and 95\% credible intervals for parameters} 
\begin{tabular}{rrrr}
  \hline
 & Mean & Lower 95\% & Upper 95\% \\ 
  \hline
$\lambda$ & 0.09 & 0.08 & 0.10 \\ 
  $\sigma^2_y$ & 0.47 & 0.41 & 0.62 \\ 
  $\sigma^2_{b_1}$ & 0.82 & 0.69 & 1.08 \\ 
  $\sigma^2_{b_2}$ & 0.28 & 0.10 & 0.41 \\ 
  $\rho_b$ & 0.41 & 0.18 & 0.77 \\ 
   \hline
\end{tabular}
\label{estimates}
\end{table}

\subsection{Distribution of Usual MVPA}

\begin{figure}[h]
\centering
\includegraphics[width=\textwidth]{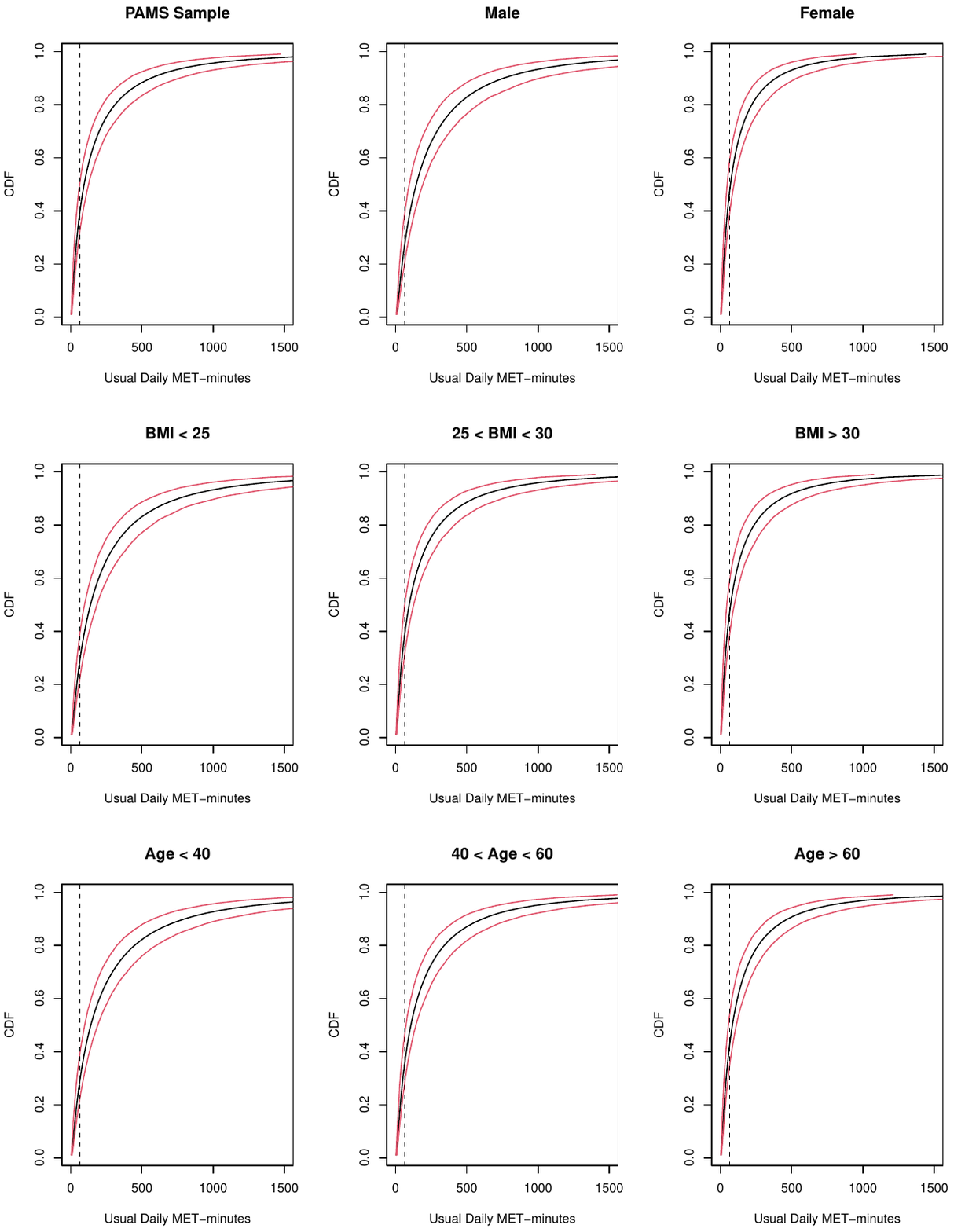}
\caption{Estimated distribution of daily MET-minutes in MVPA, with 95\% CI, for select populations using PAMS. The PAG recommendation for daily MET-minutes is the vetical dashed line.}
\label{eefigure}
\end{figure}

In Section \ref{sec:usual_mvpa} we explained how to generate distributions of MVPA in MET-minutes for any population of interest. Here we consider the PAMS population, and differences in gender, BMI, and age. Table \ref{pams_usual} shows PAG compliance rates for these different populations. The mean compliance rates and 95\% credible intervals for the PAMS sample was 0.6 (0.46, 0.69). Figure \ref{eefigure} shows the distribution of daily usual MET-minutes for each of these populations with uncertainty. Other compliance rates match what the regression coefficients in the previous section suggested, i.e. that male's tend to have higher compliance as well as younger people, and those with lower BMI. Overall, these numbers are high compared to compliance across the entire United States \cite{tucker11}. However, these results also show there is significant variability among the population, indicating that interventions targeted to specific subpopulations could be more effective than targeting the entire adult population.

\begin{table}[ht]
\centering
\caption{Estimated Physical Activity Guidelines (PAG) compliance rates, with 95\% Credible Intervals (CIs), for select populations using PAMS. BMI indicates Body Mass Index.}
\begin{tabular}{l|rr}
  \hline
 & PAG Comply & 95\% CI \\ 
  \hline
PAMS & 0.6 & (0.46,0.69) \\ 
  Male & 0.72 & (0.57,0.8) \\ 
  Female & 0.52 & (0.39,0.63) \\ 
  BMI$<$25 & 0.71 & (0.57,0.79) \\ 
  25$<$BMI$<$30 & 0.61 & (0.47,0.7) \\ 
  BMI$>$30 & 0.53 & (0.38,0.63) \\ 
  Age$<$40 & 0.7 & (0.59,0.79) \\ 
  40$<$Age$<$60 & 0.64 & (0.51,0.73) \\ 
  Age$>$60 & 0.57 & (0.43,0.68) \\ 
   \hline
\end{tabular}
\label{pams_usual}
\end{table}

\subsection{Usual MVPA using NHANES Data}

Because the results for the PAMS showed a high level of compliance in Iowa, we apply this same model to a nationally representative survey, NHANES. NHANES is a large national survey that can be used to assess the health of Americans. The 2003-2006 NHANES is the most recent collection which included physical activity monitoring with accelerometers (ActiGraph AM-7164) worn on the hip. The aim was to compare the results from the PAMS study to a different large survey that collected accelerometry information. So that results obtained from the two surveys would be comparable, we used the method proposed by \cite{ho} to select a subsample from the NHANES participants of equal size to PAMS and that would match the PAMS sample in other important ways like demographics. We implemented the method using their R package \texttt{MatchIt} \cite{ho2011}. The subsample from NHANES was selected such that each person in PAMS was matched to someone from NHANES on demographic variables including gender, age,  race, education, and BMI. Unfortunately, NHANES does not report participants' occupation, a variable that we found to be significantly associated with physical activity. For the individuals we include from NHANES, we randomly sampled two days of accelerometer measurements from the six available days. To compute bouts for the NHANES participants, we used the minute to minute information and follow the approach suggested by \cite{tucker}, and the threshold for moderate activity to be 2020 counts per minute. Counts during minutes within bouts were then converted to MET-minutes using the method of \cite{freedson}. 

\begin{figure}[h]
\centering
\includegraphics[width=\textwidth]{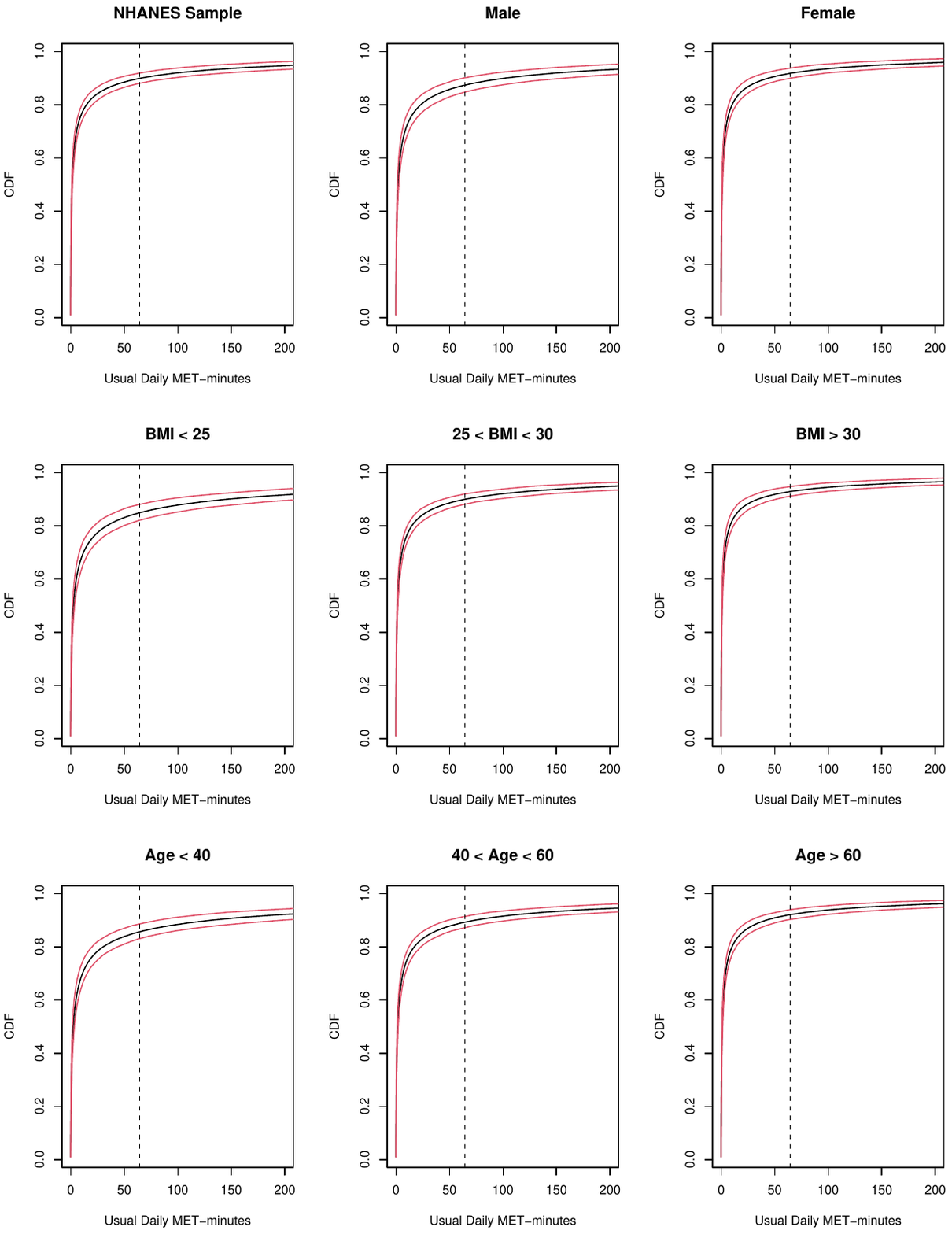}
\caption{Estimated distribution of daily MET-minutes in MVPA, with 95\% CI, for select populations using NHANES. The PAG recommendation for daily MET-minutes is the vetical dashed line.}
\label{eefigure_nhanes}
\end{figure}

The same model is fit to the subset of NHANES data. Estimated compliance with the PAG for the US population, as well as for the same populations in Table \ref{nhanes_usual}, are shown in Table \ref{nhanes_usual}. Figure \ref{eefigure_nhanes} shows the estimated distribution of daily MET-minutes for these same populations. The results for NHANES are similar to those in \cite{tucker11}, but there is a large difference when compared to the results using PAMS. Levels of activity are much lower in the NHANES data. These large differences may be attributed to several differences between PAMS and NHANES: i) PAMS is a sample of the  population of four Iowa counties while NHANES is a nationally representative sample, ii) PAMS used the SWA to measure physical activity while NHANES used the Actigraph accelerometer, iii) compliance and wear time were much higher for PAMS, iv) the SWA uses a proprietary algorithm to calculate METs while we used Freedson et al.'s method to compute METs for NHANES. Finally, over 10 years elapsed between the two surveys. Consequently, we can expect differences in terms of the desirability of participating in physical activity. 

Although the populations from which the samples were drawn are not directly comparable, we would not expect such a large difference between the two populations. Participants in PAMS wore their monitor for the entire day and night while NHANES participants were instructed to wear the device during waking hours, so this difference in wear time should not have a major effect on the measurement of MVPA. We  believe that the major differences can be at least partially attributed to the variability in different brands of accelerometers and  the way in which they convert movement to activity levels/METs. There is a large variety of methods and considerable variation between the methods of converting counts to METs \cite{crouter}.

\begin{table}[ht]
\centering
\caption{Estimated Physical Activity Guidelines (PAG) compliance rates, with 95\% Credible Intervals (CIs), for select populations using NHANES. BMI indicates Body Mass Index.} 
\label{nhanes_usual}
\begin{tabular}{l|rr}
  \hline
 & PAG Comply & 95\% CI \\ 
  \hline
NHANES & 0.1 & (0.08,0.12) \\ 
  Male & 0.12 & (0.1,0.16) \\ 
  Female & 0.08 & (0.06,0.11) \\ 
  BMI$<$25 & 0.15 & (0.12,0.19) \\ 
  25$<$BMI$<$30 & 0.1 & (0.08,0.12) \\ 
  BMI$>$30 & 0.07 & (0.05,0.09) \\ 
  Age$<$40 & 0.14 & (0.11,0.18) \\ 
  40$<$Age$<$60 & 0.11 & (0.08,0.13) \\ 
  Age$>$60 & 0.08 & (0.06,0.1) \\ 
   \hline
\end{tabular}
\end{table}

\section{Model Assessment}
\label{sec:assessment}

To assess how well our model fits the PAMS and NHANES data, we generated $M=1000$ replicate data sets from the respective posterior predictive distribution and compared selected statistics computed from the replicated datasets and from the original sample.   From these comparisons, we calculate posterior predictive p-values. Details of this procedure are in Section 5 of the supplemental material.

To assess the fit of $Y_1$, we count the number of individuals who had zero bouts on day one and zero bouts on day two, the number of individuals who had one bout on day one and zero bouts on day two, and so on for all combinations of 0,1,2+ bouts. We stop at 2+ because if an individual has two bouts in a day, they will almost certainly achieve the recommended time in MVPA. Doing this for all $M=1000$ simulated data sets, we calculate means for each category across all simulated data sets and compare to our observed proportions using a Chi-square test for proportions. We also do this procedure using a Negative Binomial distribution for $Y_1$ instead of a Generalized Poisson, swapping distributional forms in Equation \eqref{y1model}. Table \ref{y1check} shows the results. The large p-value for the Generalized Poisson here indicates that data simulated from the fitted model look similar to the observed data, at least with respect to the specific statistic. The small p-value related to the Negative Binomial model for $Y_1$ indicates a lack of fit, and therefore the Generalized Poisson model is preferred in this application. 

\begin{table}
\centering
\caption{Chi-square test for proportions comparing the mean values from simulated data sets to observed values for $Y_1$ using both a Generalized Poisson distribution and Negative Binomial.  Results for Generalized Poisson: $\chi^2$ = 3.7705, df=8, p-value = 0.8772. Results for Negative Binomial: $\chi^2$ = 82.313, df=8, p-value $<$ 1e-6.}
\begin{tabular}{ll|lll}
\hline
Number of & bouts & Observed & Generalized Poisson & Negative Binomial \\
\hline
0 &0 & 126 & 132 & 158\\
1 &0 & 57 & 65 & 42\\
2+& 0 & 77 & 78 & 142\\
0& 1 & 65 & 65 & 41\\
0& 2+ & 71 & 78 & 141\\
1 &1 & 48 & 46 & 15\\
1& 2+ & 81 & 83 & 62\\
2+& 1 & 91 & 83 & 62\\
2+& 2+ & 441 & 424 & 392\\
\hline
\end{tabular}

\label{y1check}
\end{table}

We also calculate the mean within-person standard deviation of $Y_1$ and the within-person range of $Y_1$. 
The posterior predictive p-values for these are 0.7 and 0.139, respectively, which indicates no lack of fit.

To assess the overall fit of the non-zero values of $Y_2$, we use the Kolmogorov-Smirnov test to compare each simulated data sets' empirical cumulative distribution function (ecdf) from the fitted model to the observed values' ecdf of $Y_2$. We perform this test for all $M$ simulated data sets, so we have $M$ p-values. Table \ref{y2check} shows a summary of those p-values. These results show there are not apparent issues in the fit of $Y_2$ either. We performed the same model assessment procedures after fitting the model to the NHANES data, and the results were similar, indicating that the model also appears to fit the NHANES data well.

\begin{table}
\centering
\caption{Summary of $M=1000$ Kolmogorov-Smirnov test p-values comparing simulated data sets from fitted model to the observed data.}
\begin{tabular}{l|l|l|l}
\hline
1st Quantile & 2nd Quantile & 3rd Quantile & Mean \\
\hline
0.129 & 0.377 & 0.693 & 0.417 \\
\hline
\end{tabular}

\label{y2check}
\end{table}

\section{Discussion}

This paper presented a two-part Bayesian hierarchical model with measurement error that can be used to estimate MET-minutes in MVPA. In turn, the model can further be used to estimate compliance with the PAG. We were able to accommodate the recommendation that activity must come in at least 10 minute bouts by jointly modeling the number of bouts and average excess MET-minutes per bout for individuals. Additionally, these were modeled as functions of demographic variables which could then be used to create distributions for subpopulations. We used data from the PAMS study to fit the model. In PAMS, participants wore an activity monitor on two separate days, for 24 hours.  In preliminary analysis, we found that the 2-3 week buffer between measurements in  PAMS  seemed to successfully remove any dependence between recording days. The results showed that men and those with jobs that are physically demanding had higher levels of MVPA, and those with college degrees did as well but to a lesser extent. Age and BMI were negatively associated with MVPA. This type of information might be useful in designing interventions and that target specific subpopulations.

The estimated distributions of usual MVPA that were based on the PAMS data were unexpected in that about 60\% of the Iowa adult population met the current the PAG.  The high proportion of compliers is at odds with the rates of obesity and the sedentary lifestyle that have been documented \cite{owen}. 
Based on the raw data, only 27\% of the sample didn't achieve sufficient condition of two bouts per day to meet PAG guidelines. Moreover, only 11\% didn't participate in a bout of MVPA. There are various interpretations for these results. First, there are differences in reported activity when accelerometers are worn on the hip versus the wrist or arm. Both \cite{rosenberger} and \cite{ellis} found higher accuracy when accelerometers were worn on the hip. Since the SWA is worn on the arm, it can capture upper body activity and potentially record it as MVPA when it is not. This is one argument for why the results from PAMS seem so high. In addition to these problems, the SWA is known to overestimate MVPA \cite{scheers, santos}. Second, it is possible that the PAG are set at a level that is easy to meet and that health benefits are realized with higher levels of physical activity. In contrast, the results we obtained using NHANES data suggest that only 10\% of American adults are in compliance with the PAG. 

  New methods that do not assume unbiasedness of accelerometry measurements are needed. To fit these new models, we require a gold standard to measure minute by minute physical activity in order to calibrate accelerometry measurements. To further complicate things, \cite{hills} claims that it is ``unlikely that a single measure of reported PA would suffice", in reference to assessing every possible activity in which humans engage. Finally, the PAG also advises adults to participate in two sessions of muscle building activity per week to realize health effects. The PAMS did not measure this type of physical activity, and thus we did not consider it in our calculation of compliance rates.

\section*{Acknowledgements}

The authors gratefully acknowledge the insightful suggestions of Professor Wayne F. Fuller. We are also grateful to Drs. David Osthus and Bryan Stanfill, who were responsible for collecting and assembling the data we used in this work. We also thank Dr. Kevin W. Dodd from NCI with whom we had fruitful discussions and who generously shared code for some of the initial processing of the NHANES data. Finally, we thank Elizabeth Schneider for her help proofing and help getting this manuscript in final form.

\section*{Funding}
This work was supported by National Institutes of Health Grant number HL091024. No potential conflict of interest was reported by the authors. Sandia National Laboratories is a multimission laboratory managed and operated by National Technology and Engineering Solutions of Sandia, LLC, a wholly owned subsidiary of Honeywell International Inc., for the U.S. Department of Energy's National Nuclear Security Administration under contract DE-NA0003525. This paper describes objective technical results and analysis. Any subjective views or opinions that might be expressed in the paper do not necessarily represent the views of the U.S. Department of Energy or the United States Government.

\section*{Data Availability Statement}

The data that support the findings of this study are available from the corresponding author, DR, upon reasonable request. NHANES data is available at https://wwwn.cdc.gov/nchs/nhanes/default.aspx.

\bibliographystyle{tfs}
\bibliography{mybib}

\clearpage

\section*{Supplemental Material}

\subsection*{Example of raw data given by PAMS SWA}

\begin{figure}[h]
\centering
\includegraphics[width=.8\textwidth]{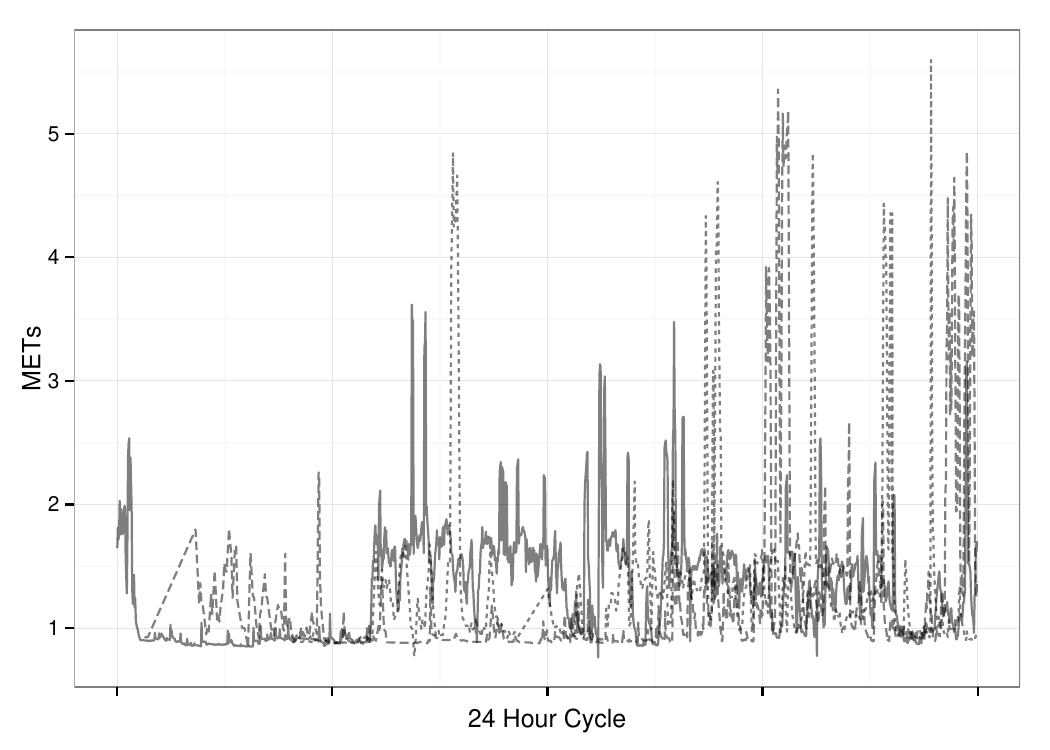}
\caption{24 hour MET plot for three individuals in PAMS (given by different line types) over 24 hours.}
\label{rawdata}
\end{figure}

Figure \ref{rawdata} gives an example of three individuals' plot of MET activity across 24 hours. MET levels are often hovering around 1.5 during waking hours with a couple short duration spikes in MET activity during the day. 

\subsection*{Distribution of Average Excess MET-minutes $Y_2$}
Using log-transformed the positive $Y_2$ values, we performed the linear regression:

\begin{align}
	log (Y_{2ij}) &= \boldsymbol{\theta}'{\bf Z}_i + e_{ij},\\
	e_{ij} &\overset{iid}{\sim} N(0,\sigma^2).
\end{align}

\begin{figure}[h]
\centering
\includegraphics[width=3.5in, height=2.5in]{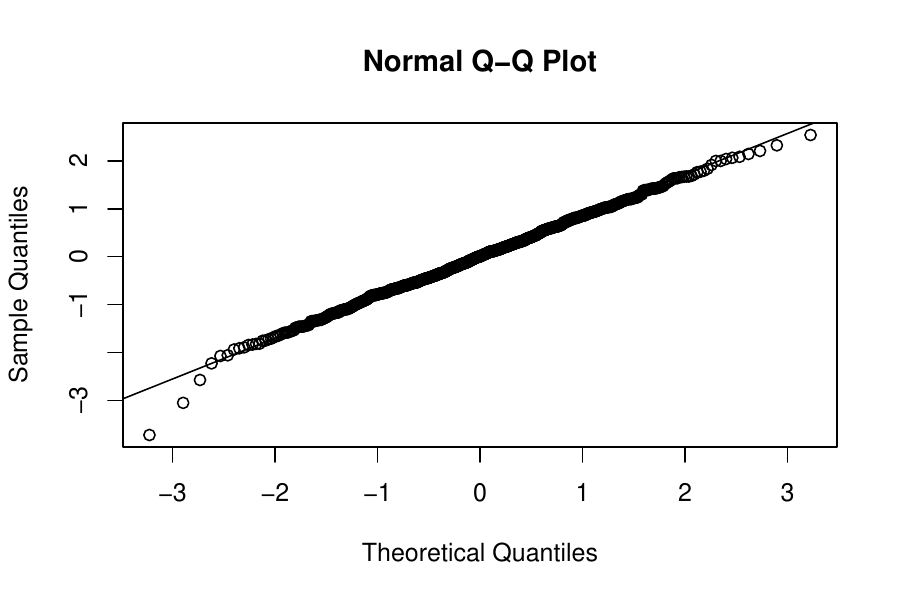}
\caption{Normal quantile plot of residuals for log transformed $Y_2$ regression. This plot shows a log($Y_2$) is approximately Normally distributed.}
\label{qqplot3}
\end{figure}

Figure \ref{qqplot3} shows a QQ plot for the residuals for the above model. A Shapiro-Wilk test for normality of the residuals results in a p-value of 0.15; this along with the QQ plot suggests that the empirical distribution of the log transformed data approximates a normal distribution, which allows us to use a lognormal distribution to model the $Y_2$ in the original scale.

\subsection*{Bowker's Test to justify Exchangeability Assumption}

\begin{figure}[h]
\includegraphics[height=3in,width=5in]{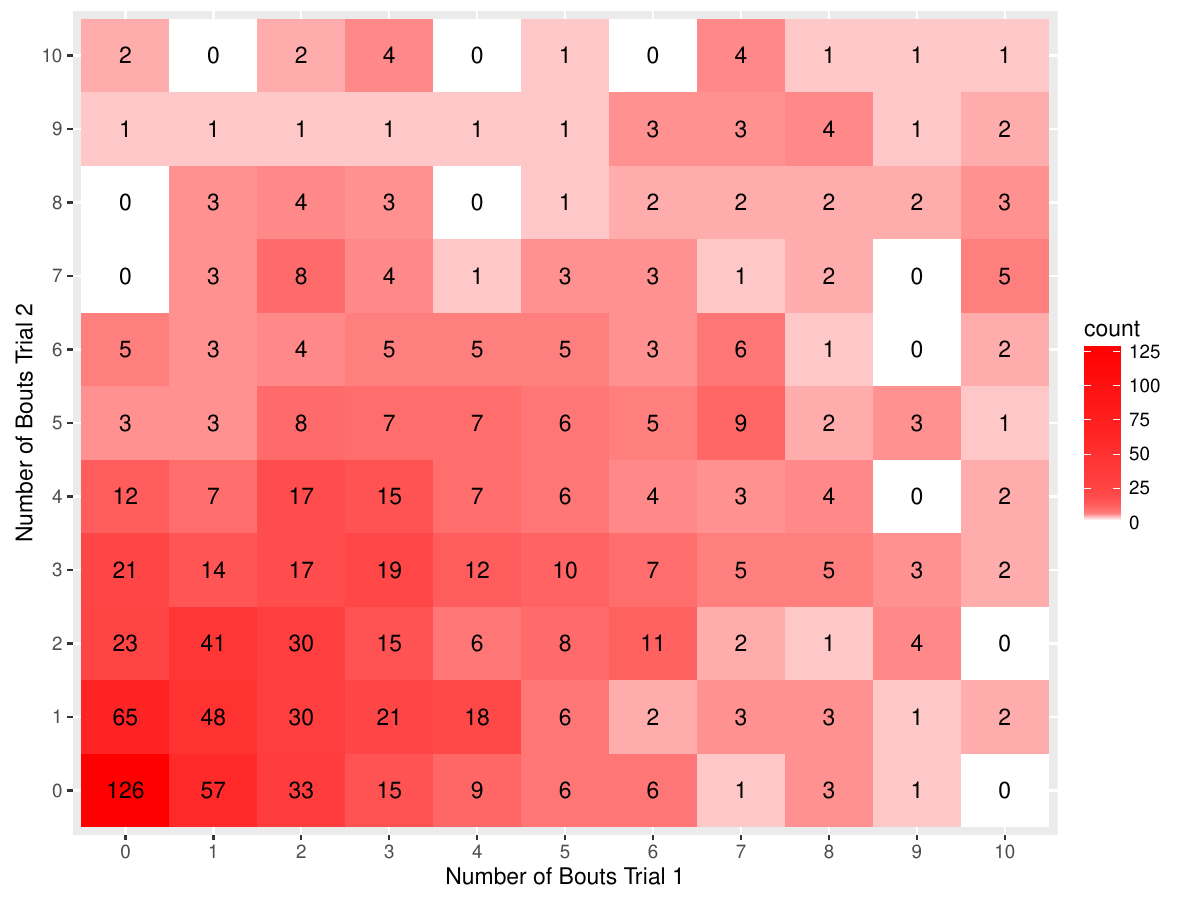}
\caption{2 Contingency way table for individuals with two observations based on number of bouts per trial. Truncated at 10 bouts due to sparsity beyond that.}
\label{y1exch}
\end{figure}

To check whether assuming observations within an individual for $Y_1$ are exchangeable is reasonable,  including only individuals who had two observations. Figure \ref{y1exch} shows the frequency of individuals that had the particular combination of bouts on days one and two.  Bowker proposed a test for symmetry in $m$ by $m$ contingency tables. The null hypothesis of Bowker's test is that $\pi_{lk}=\pi_{kl}$ $\forall$ $l \neq k$ where $\pi_{ij}$ is the true frequency in the $ij$th cell. 
We tested the symmetry of the contingency table 
and the p-value was 0.12. This test is sensitive to the presence of zero or low counts, so we also implemented the same test on a smaller subset of the contingency table (number of bouts up to 6) to ensure that the results were consistent. In all cases, we failed to reject the null hypothesis, which suggests that within individual measurements of number of bouts can be assumed to be exchangeable. 

\subsection*{Within-person Overdispersion of $Y_1$}

\begin{figure}[h]
\centering
\includegraphics[width=0.6\textwidth]{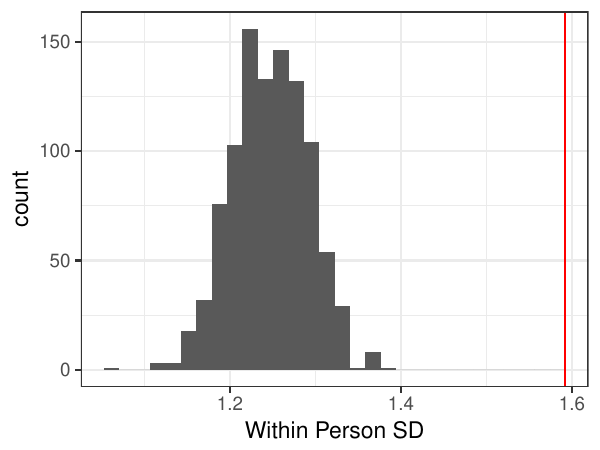}
\caption{Posterior predictive model assessment for Poisson model. Statistic is mean within person standard deviation. Vertical line indicates observed value.}
\label{badmodel}
\end{figure}

We fit a Poisson model to the $Y_1$ data, and simulated data from the fitted model. Figure \ref{badmodel}  shows the distribution of mean within person standard deviations for each simulated dataset as well as the truth as a vertical line. This shows the standard Poisson distribution is not sufficient for these data.

\clearpage
\subsection*{Full conditional distributions}

\begin{align}
\gamma|\cdot &\propto \left[  \prod_{i=1}^{1057} \prod_{j=1}^{2} f(Y_{1ij}| Z_i, b_{1i},\gamma,\lambda) f(Y_{2ij}|Z_i,b_{2i},\beta,\sigma_y^2,\pi_i) \right] p(\gamma) \\
&= \left[  \prod_{i=1}^{1057} \prod_{j=1}^{2} \mu_1(1-\lambda)(\mu_1(1-\lambda)+Y_{1ij}\lambda)^{Y_{1ij}-1} e^{-\mu_1(1-\lambda)-Y_{1ij}\lambda} \frac{1}{Y_{2ij}\sigma_y \sqrt{2\pi}} e^{-\frac{1}{2\sigma_y^2}(ln Y_{2ij}-Z_i`\beta - b_2i)^2} \right] \\
&\times e^{-\frac{1}{200}\gamma^2} \\
&\text{where } \mu_1 = e^{Z'\gamma + b_{1i}} \\
\beta|\cdot &\propto \left[  \prod_{i=1}^{1057} \prod_{j=1}^{2}  f(Y_{2ij}|Z_i,b_{2i},\beta,\sigma_y^2,\pi_i)I(Y_{2ij}>0) \right] p(\beta) \\
&\sim N(m_{\beta},V_{\beta}) \\
&V_{\beta} = (V_0^{-1} + \frac{1}{\sigma_y^2} Z'Z)^{-1}, m_{\beta} = V_{\beta}(V_0^{-1}m_0 + Z'(log {\bf Y_2 - b_2})/\sigma_y^2), \text{ for $Y_{2ij} > 0$} \\
\lambda|\cdot &\propto \left[  \prod_{i=1}^{1057} \prod_{j=1}^{2} f(Y_{1ij}| Z_i, b_{1i},\gamma,\lambda) f(Y_{2ij}|Z_i,b_{2i},\beta,\sigma_y^2,\pi_i) \right] p(\lambda) \\
&= \left[  \prod_{i=1}^{1057} \prod_{j=1}^{2} \mu_1(1-\lambda)(\mu_1(1-\lambda)+Y_{1ij}\lambda)^{Y_{1ij}-1} e^{-\mu_1(1-\lambda)-Y_{1ij}\lambda} \frac{1}{Y_{2ij}\sigma_y \sqrt{2\pi}} e^{-\frac{1}{2\sigma_y^2}(ln Y_{2ij}-Z_i`\beta - b_2i)^2} \right] \\
& \times  I(0<\lambda < 1)
\end{align}
\begin{align}
\sigma_y^2|\cdot &\propto \left[  \prod_{i=1}^{1057} \prod_{j=1}^{2}  f(Y_{2ij}|Z_i,b_{2i},\beta,\sigma_y^2,\pi_i)I(Y_{2ij}>0) \right] p(\sigma_y^2) \\
&\sim \text{Inverse-Gamma}\left( \frac{N^*}{2} + a_0, \frac{1}{2} \sum_{i=1}^n \sum_{j=1}^{2} (log Y_{2ij} - Z_i' \beta - b_{2i})^2 \right) \\
&\text{ for $Y_{2ij} > 0$ where $N^*$ is the number of non-zero observations of $Y_{2ij}$}\\
\Sigma_b|\cdot &\propto \left[ \prod_{i=1}^{n} f(b_{1i},b_{2i}|\Sigma_b) \right] p(\Sigma_b) \\ 
&\sim \text{Inverse-Wishart}(n+d_0, {\bf b'b} + D_0) \\
&\text{ where {\bf b} is an $n\times 2$ matrix with the first column containing elements of $b_1$, } \\
&\text{and second column containing elements of $b_2$}  \\
b_{1i},b_{2i}|\cdot &\propto \left[  \prod_{j=1}^{2} f(Y_{1ij}| Z_i, b_{1i},\gamma,\lambda) f(Y_{2ij}|Z_i,b_{2i},\beta,\sigma_y^2,\pi_i) \right] f(b_{1i},b_{2i}|\Sigma_b) \\
&=  \left[  \prod_{j=1}^{2} \mu_1(1-\lambda)(\mu_1(1-\lambda)+Y_{1ij}\lambda)^{Y_{1ij}-1} e^{-\mu_1(1-\lambda)-Y_{1ij}\lambda} \frac{1}{Y_{2ij}\sigma_y \sqrt{2\pi}} e^{-\frac{1}{2\sigma_y^2}(ln Y_{2ij}-Z_i`\beta - b_2i)^2} \right] \\
&\times e^{-\frac{1}{2} {\bf b_i}' \Sigma_b^{-1} {\bf b_i}} 
\end{align}

\clearpage
\subsection*{Posterior Predictive p-value}

Let $\boldsymbol{\theta}$ represent all unknown parameters except  the latent variables.  To simulate a new data set, we generate $M$ random draws from the posterior distribution $p(\boldsymbol{\theta}|{\bf Y}_1,{\bf Y}_2, {\bf Z})$. We use these draws as well as the  values of the covariates to simulate new latent variables  ${\bf b}_1,{\bf b}_2$ from $p({\bf b}_1,{\bf b}_2|{\bf Z},\boldsymbol{\theta_2}) $.  Finally, with the latent variables and  posterior draws, we simulate new observations  ${\bf Y}_1$ from the data model $p({\bf Y}_1^*|{\bf b}_1,{\bf Z},\boldsymbol{\theta}_1)$ and ${\bf Y}_2$ from $ p({\bf Y}_2^*|{\bf b}_1,{\bf b}_2,{\bf Z},\boldsymbol{\theta}_1) $. A posterior predictive p-value for statistic $T(Y,\boldsymbol{\theta})$ is calculated by:

\begin{align}
p-value &= \frac{1}{M} \sum_{i=m}^{M} I(T(Y_m^*,\boldsymbol{\theta}_m) < T(Y^{obs},\boldsymbol{\theta})).
\end{align}

\subsection*{Prior Sensitivity Analysis}

In addition to the priors selected in the main paper, we also considered other priors in a sensitivity analysis.

\subsubsection*{Prior set 2}
\begin{align}
	\lambda &\sim \text{Uniform}(0,1) \\
	\boldsymbol{\gamma} &\sim N \left( {\bf 0}_8, 1000 I_{8 \times 8} \right) \\
	\Sigma_b &\sim \text{Inverse-Wishart}(8, 5*I_{2 \times 2}) \\
	\sigma_y^2 &\sim \text{Inverse-Gamma}(5,5), \\
	\boldsymbol{\beta} &\sim N \left( {\bf 0}_8, 1000 I_{8 \times 8} \right)
\end{align}

\subsubsection*{Prior set 3}
\begin{align}
	\lambda &\sim \text{Uniform}(0,1) \\
	\boldsymbol{\gamma} &\sim N \left( {\bf 0}_8, 1 I_{8 \times 8} \right) \\
	\Sigma_b &\sim \text{Inverse-Wishart}(4, .1*I_{2 \times 2}) \\
	\sigma_y^2 &\sim \text{Inverse-Gamma}(.1,.1), \\
	\boldsymbol{\beta} &\sim N \left( {\bf 0}_8, 1 I_{8 \times 8} \right)
\end{align}

\subsubsection*{Prior set 4}
\begin{align}
	\lambda &\sim \text{Uniform}(0,1) \\
	\boldsymbol{\gamma} &\sim N \left( {\bf 5}_8, 100 I_{8 \times 8} \right) \\
	\Sigma_b &\sim \text{Inverse-Wishart}(4, I_{2 \times 2}) \\
	\sigma_y^2 &\sim \text{Inverse-Gamma}(.1,.1), \\
	\boldsymbol{\beta} &\sim N \left( {\bf 5}_8, 100 I_{8 \times 8} \right)
\end{align}

\begin{figure}[h]
\includegraphics[width=\textwidth]{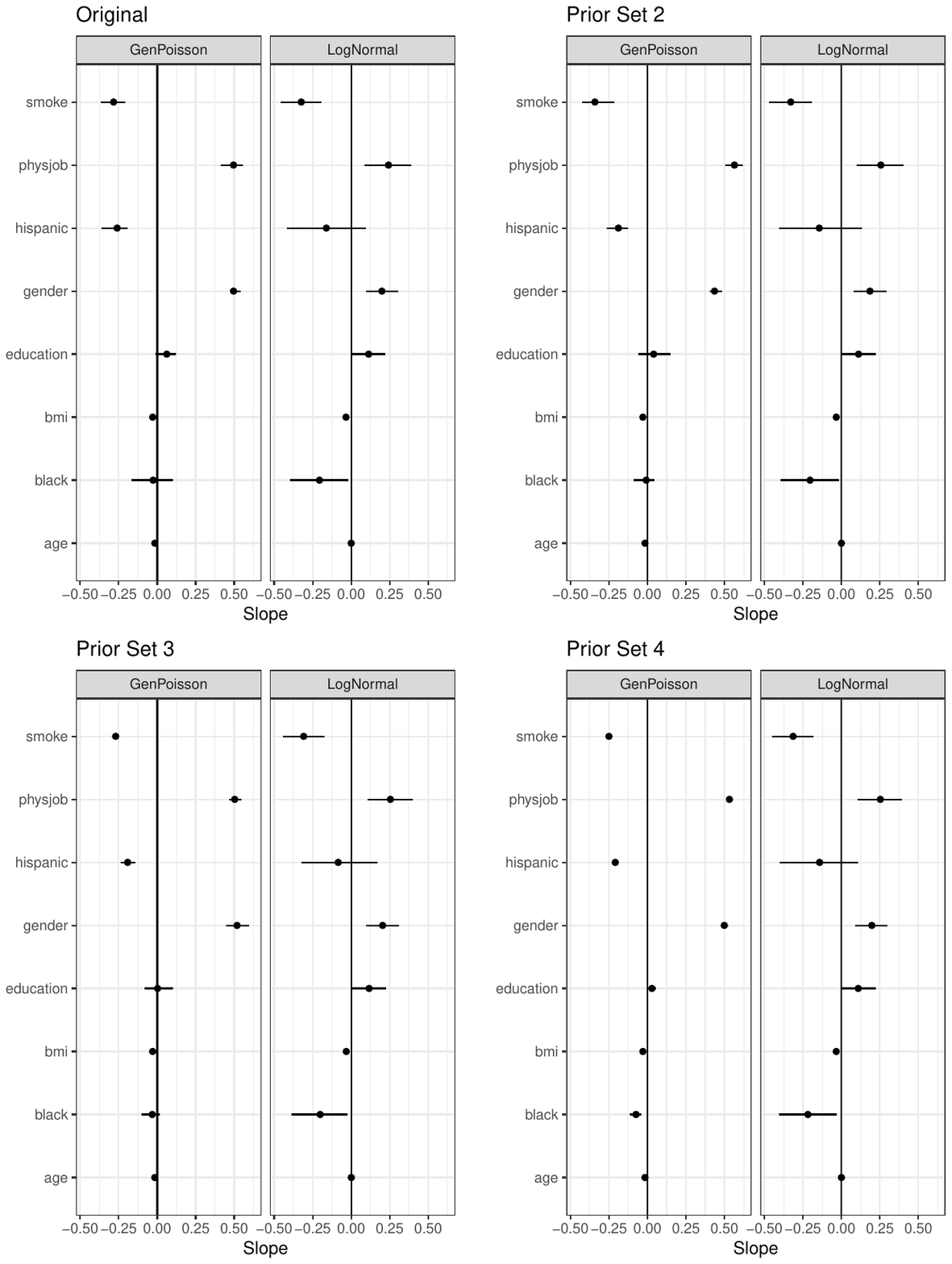}
\caption{Regression coefficients posterior medians and 95\% CI for different sets of priors.}
\label{prior}
\end{figure}

Figure \ref{prior} shows the posterior medians and 95\% CIs for the $\gamma$ and $\beta$ regression coefficients. There are only small differences between the different prior sets. 

\end{document}